\documentclass[fleqn,usenatbib]{mnras}

\usepackage{newtxtext,newtxmath}

\usepackage{listings}
\usepackage[T1]{fontenc}
\usepackage{ae,aecompl}

\usepackage{graphicx}	% Including figure files
\usepackage{amsmath}	% Advanced maths commands
\usepackage{amssymb}	% Extra maths symbols

\defcitealias{sh16}{SH16}

\title[Inference using noisy simulations]{Parameter inference and model comparison using theoretical predictions from noisy simulations}

\author[N. Jeffrey and F. B. Abdalla]{
Niall Jeffrey$^{1}$\thanks{E-mail: niall.jeffrey.15@ucl.ac.uk}
and Filipe B. Abdalla$^{1,2}$
\\
$^{1}$ Department of Physics \& Astronomy, University College London, Gower Street, London, WC1E 6BT, UK\\
$^{2}$ Department of Physics and Electronics, Rhodes University, PO Box 94, Grahamstown, 6140, South Africa\\
}

\date{Accepted 2019 September 30. Received 2019 September 26; in original form 2018 September 13.}

\pubyear{2018}

\begin{document}
\label{firstpage}
\pagerange{\pageref{firstpage}--\pageref{lastpage}}
\maketitle

\begin{abstract}
When inferring unknown parameters or comparing different models, data must be compared to underlying theory. Even if a model has no closed-form solution to derive summary statistics, it is often still possible to simulate mock data in order to generate theoretical predictions. For realistic simulations of noisy data, this is identical to drawing realizations of the data from a likelihood distribution. Though the estimated summary statistic from simulated data vectors may be unbiased, the estimator has variance which should be accounted for. We show how to correct the likelihood in the presence of an estimated summary statistic by marginalizing over the true summary statistic in the framework of a Bayesian hierarchical model. For Gaussian likelihoods where the covariance must also be estimated from simulations, we present an alteration to the Sellentin-Heavens corrected likelihood. We show that excluding the proposed correction leads to an incorrect estimate of the Bayesian evidence with JLA data. The correction is highly relevant for cosmological inference that relies on simulated data for theory (e.g. weak lensing peak statistics and simulated power spectra) and can reduce the number of simulations required.
\end{abstract}

\begin{keywords}
methods: statistical -- methods: data analysis -- cosmology: observations
\end{keywords}

%%%%%%%%%%%%%%%%%%%%%%%%%%%%%%%%%%%%%%%%%%%%%%%%%%

%%%%%%%%%%%%%%%%% BODY OF PAPER %%%%%%%%%%%%%%%%%%

\section{Introduction}

It is increasingly common, especially in cosmological surveys, to attempt to make inferences from data $\mathbf{d}$ using theory summary statistics $\mathbf{\mu}$ that can be obtained only from simulations. 

One example, currently popular in cosmology, is weak lensing peak statistics (\citealt{peaks2010};~\citealt{des_peaks};~\citealt{peel_peaks};~\citealt{kids_peaks1};~\citealt{kids_peaks2}). Peak statistics broadly aim to use the number of density peaks in the cosmological matter distribution to constrain cosmological parameters and models. The number of peaks in the density field (or weak lensing signal) is the result of highly non-linear structure formation and large-volume dark matter simulation are often needed to generate the theoretical number. The number of peaks in a given simulation is stochastic due to cosmic variance and then further sources of noise are added to simulate realistic observed data.

The data $\mathbf{d}$ are not the raw maps or catalogues, but the observed/derived summary statistics. For example, $\mathbf{d}$ would be the observed number of peaks in a weak lensing mass map reconstruction (\citealt{des_peaks};~\citealt{des_mass_map_peaks}).

It was recently noted by \cite{sh16} that the common practice of debiasing a covariance matrix estimated from simulations of mock data~\citep{hartlap06} is insufficient. The sampling distribution of this estimated covariance should be incorporated into the likelihood distribution and, therefore, into the posterior distributions of the inferred parameters. Failure to do so leads to biased and overly-optimistic inferences.

In this work we note that, as with the estimated covariance described by \cite{sh16}, an unbiased estimated summary statistic $\hat{\mathbf{\mu}}$ is nevertheless itself a random variable, drawn from a sampling distribution with associated variance. If unaccounted for, this will lead to inaccurate parameter inference and misleading model comparison results. 

In a Bayesian hierarchical framework, we present how to derive the posterior distribution of parameters by using a corrected likelihood distribution which takes into account that the theoretical predictions are based on noisy simulations. 

An alternative to our presented hierarchical approach is to use likelihood-free inference (LFI) methods (e.g.~\citealt{alsing_likelihood_free},~\citealt{leclercq_likelihood_free}). In general LFI methods assume that the likelihood is unknown, and simulations are used to estimate the resulting posterior distribution conditional on data. However, if simulations are expensive and we believe we know the naive likelihood\footnote{Assuming we could condition on the true summary statistics which can be calculated $P(\mathbf{d} | \mathbf{\mu}(\mathbf{\theta})).$} then LFI methods would be unnecessary and require more simulations, due to the increased number of degrees of freedom in the model. Instead, we show how to directly construct a Bayesian hierarchical model containing the free parameters of the known naive likelihood.

In Section~\ref{sec:correction} we show how to marginalize over possible summary statistics $\mathbf{\mu}$ to derive the likelihood $P(\mathbf {d}\ |\ \hat{\mathbf{\mu}})$, the probability of the data  $\mathbf{d}$ conditional on the estimated summary statistic $\hat{\mathbf{\mu}}$. In Section~\ref{sec:gaussian}, we consider the case in which the naive likelihood is Gaussian, and derive the corrected likelihood distribution in the presence of both known and unknown (estimated) covariance matrices. In Section~\ref{sec:toy_model}, we use a one-dimensional toy model to demonstrate the effect of estimated summary statistics; we show that the corrected likelihood distribution matches samples generated from the toy model. In Section~\ref{sec:JLA}, we use the public JLA supernovae data to show the effect of estimating summary statistics from simulations, using draws from the known likelihood as mock simulations.

\section{Likelihood-based inference with simulations} \label{sec:correction}

\subsection{Posterior and likelihood}

Using the example of weak lensing peak statistics, we would wish to evaluate the posterior distribution of parameters of interest $\mathbf{\theta}$ (e.g. $\Omega_m$, $\sigma_8$ etc.) conditional on our observed data $\mathbf{d}$ (the number of peaks in a weak lensing map), which is given by Bayes' theorem:

\begin{equation} \label{eq:simple_bayes}
 P(\mathbf{\theta} | \mathbf{d}) = \frac{P(\mathbf{d} |\mathbf{\theta})\ P(\mathbf{\theta}) }{P(\mathbf{d})} \ .
\end{equation}

\noindent If we were able to directly model the summary statistic with a deterministic function $\mathbf{\mu}(\mathbf{\theta})$ (e.g. if we could calculate the number of weak lensing peaks for given parameters $\mathbf{\theta}$), then we could substitute\footnote{For a deterministic $\mathbf{\mu} = \mathbf{f}(\mathbf{\theta})$, then $P(\mathbf{\mu} |\mathbf{\theta}) = \delta^{\mathrm{Dirac}} \big( \mathbf{f}(\mathbf{\theta}) - \mathbf{\mu} \big)$. Substitution is performed via marginalization over a latent $\mathbf{\mu}'$.} $\mathbf{\mu}(\mathbf{\theta})$ into the likelihood: 

\begin{equation} \label{simple_bayes}
P(\mathbf{d} |\mathbf{\theta}) = P(\ \mathbf{d}\ |\  \mathbf{\mu}(\mathbf{\theta})\ ) \ .
\end{equation} 

\noindent Of course, this is not possible in many cases where the deterministic model is not known. We must rely on estimates $\hat{\mathbf{\mu}}_i$ of the true, but unattainable, summary statistic, with simulations run at position $i$ in parameter space with parameters $\theta_i$. 

The correct representation of Bayes' theorem for the posterior distribution is

\begin{equation} \label{eq:correct_bayes}
 P(\mathbf{\theta} | \mathbf{d}, \{\hat{\mathbf{\mu}}_i \}) = \frac{P(\mathbf{d} |\{\hat{\mathbf{\mu}}_i\}, \mathbf{\theta})\ P(\mathbf{\theta}) }{P(\mathbf{d})}  \ \ ,
\end{equation}

\noindent or equivalently

\begin{equation} \label{eq:correct_bayes2}
 P(\mathbf{\theta} | \mathbf{d}, \{\hat{\mathbf{\mu}}_i \}) = \frac{P(\mathbf{d}, \{\hat{\mathbf{\mu}}_i\}  |   \mathbf{\theta})\ P(\mathbf{\theta}) }{ P(\{\hat{\mathbf{\mu}}_i\}) \ P(\mathbf{d})}  \ \ ,
\end{equation}

\noindent keeping in mind that, as observed quantities, the observed summary statistic $\mathbf{d}$ and the statistics from the simulations  $\{\hat{\mathbf{\mu}}_i \}$ are independent and, therefore, separable. The brackets $\{ \}$ represent the set of simulations run over the positions $i$ in parameter space.

The factors $P(\{\hat{\mathbf{\mu}}_i\})$ and  $P(\mathbf{d})$ are both Bayesian evidence terms for the observed data and observed simulations. As constants, they can be largely ignored.

\begin{figure}
\hspace*{-0.2cm}
\includegraphics[width=0.48\textwidth]{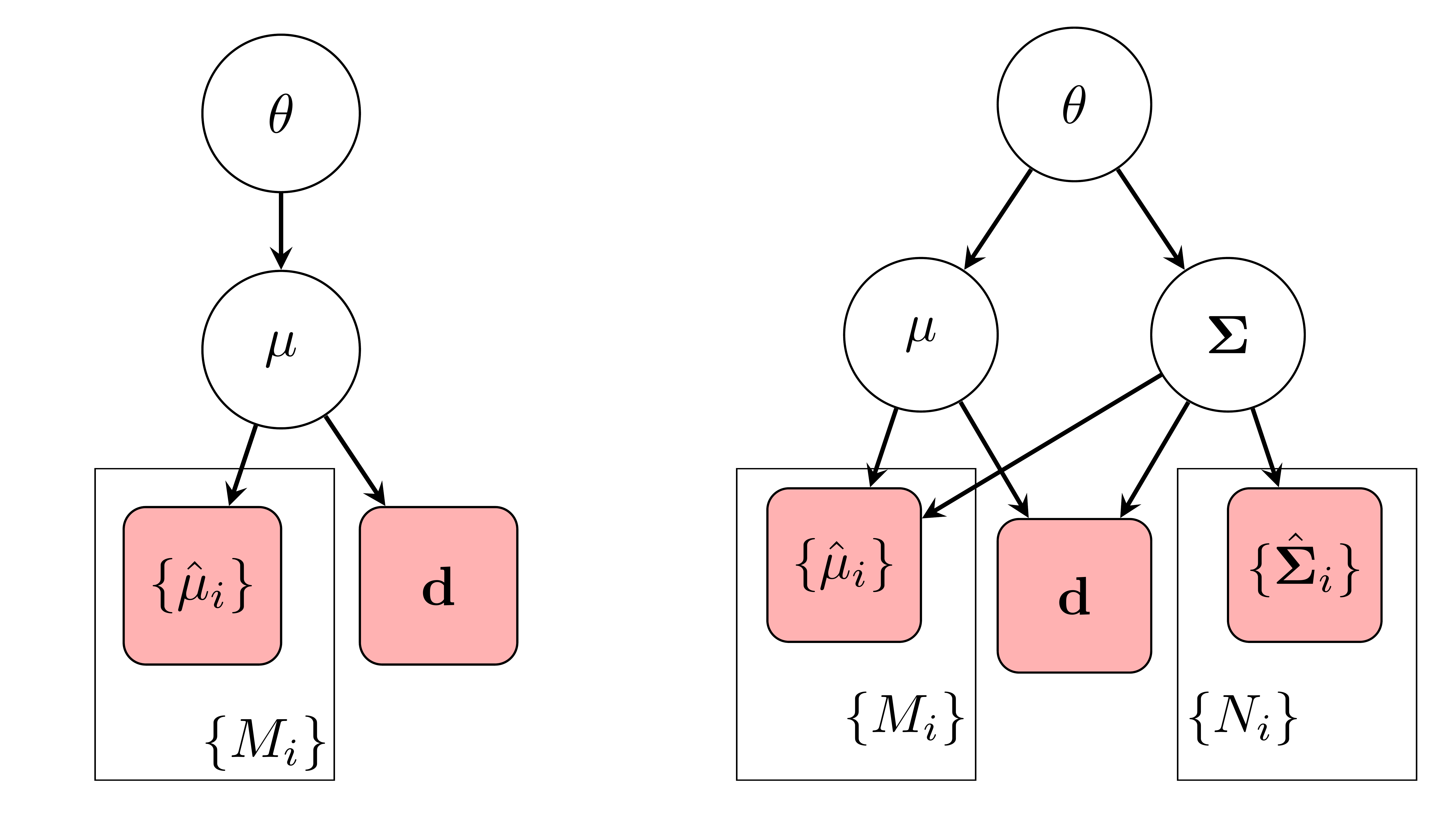}
\vspace{-.2cm}
\caption{\label{fig:network} Probabilistic graphical representation in plate notation of Bayesian hierarchical models from Sections~\ref{sec:likelihood_correction} \&~\ref{sec:known} ({\textit{left panel}}) and Section~\ref{sec:unknown}. (\textit{right panel}). Shaded nodes are ``observed', either from experimental data ($\mathbf{d}$) or from simulations ($\hat{\mathbf{\mu}}$ and $\hat{\mathbf{\Sigma}}$). Each plate (rectangular box) includes the amount of data associated with the variable, for example each $\hat{\mathbf{\mu}}_i$ (run at position $i$ in parameter space)  comes from $M_i$ simulations.}
% \vspace{-.1cm} 
\end{figure} 

\subsection{Likelihood correction}\label{sec:likelihood_correction}

At position $i$ in parameter space a set of $M$ simulated data are generated. As it is not possible to calculate\footnote{that is, $\mathbf{\mu}(\mathbf{\theta})$ cannot be calculated for a given $\mathbf{\theta}$ (e.g. peak statistics)} the summary statistic $\mathbf{\mu}$, we estimate it from $M$ simulated data realizations $\mathbf{d}_{\rm sim}$. The estimate is often the mean 

\begin{equation} \label{eq:mean_estimate}
\hat{\mathbf{\mu}} (\mathbf{\theta}_i) = \frac{1}{M} \sum_{j=0}^{M - 1} \mathbf{d}_{{\rm sim},j} (\mathbf{\theta}_i) \ \ ,
\end{equation}

\noindent where each $j_{\rm{th}}$ data realization is independent. In some cases, the summary statistic may not be the expectation of the likelihood $P(\mathbf{d} | \mathbf{\mu})$, so the estimator $\hat{\mathbf{\mu}}$ would be another function of the simulated data (not the mean).

Even if $\hat{\mathbf{\mu}}$ is an unbiased estimate (that is, $\langle \hat{\mathbf{\mu}}\rangle  =  \mathbf{\mu} $), it is often mistakenly assumed that 

\begin{equation}
P(\ \mathbf{d}\  |\ \hat{\mathbf{\mu}}_i\ ) = P(\ \mathbf{d}\ |\  \mathbf{\mu}(\mathbf{\theta}_i)\ ) \ \ ,
\end{equation} 

\noindent which is not generally correct. 

% In the scenario described, one cannot know $P(\mathbf {d}\ |\ \mathbf{\mu})$, as $\mathbf{\mu}$ itself is an unknown random variable.

% The correct likelihood to be used for parameter inference is 
% \begin{equation}
% P(\mathbf {d}\ |\ \hat{\mathbf{\mu}},\ M, \ \alpha) \ ,
% \end{equation}

% \noindent which can be rewritten as a marginalization over the true summary statistic

The correct likelihood to be used for parameter inference is with the estimated $\{\hat{\mathbf{\mu}}_i\}$ from simulations as used in equation~\ref{eq:correct_bayes} or  equation~\ref{eq:correct_bayes2}. This can be rewritten as a marginalization over the unknown true summary statistic 

\begin{equation} \label{eq:marginal}
\begin{split}
P(\mathbf {d}\ |\ \{\hat{\mathbf{\mu}}_i\} , \ \mathbf{\theta} ) &= \int P(\mathbf{d}\ ,\ \mathbf{\mu}\ |\ \{\hat{\mathbf{\mu}}_i\} , \ \mathbf{\theta}) \  \rm{d}\mathbf{\mu}  \\
&= \int P(\mathbf{d}\ | \mathbf{\mu}, \ \mathbf{\theta})  \ P(\mathbf{\mu}\ |\ \{\hat{\mathbf{\mu}}_i\} , \ \mathbf{\theta}) \  \rm{d}\mathbf{\mu}   \\
&= \frac{1}{P(\{\hat{\mathbf{\mu}}_i\})}\int P(\mathbf{d}\ | \mathbf{\mu}, \ \mathbf{\theta} )  \ P( \{\hat{\mathbf{\mu}}_i\} |\ \mathbf{\mu}, \ \mathbf{\theta}) \  P(\mathbf{\mu}| \mathbf{\theta}) \  \rm{d}\mathbf{\mu}    \ ,
\end{split}
\end{equation}

\noindent or alternatively

\begin{equation} \label{eq:marginal2}
P(\mathbf {d}, \{\hat{\mathbf{\mu}}_i\} |\ \ \mathbf{\theta} ) = \int P(\mathbf{d}\ | \mathbf{\mu}, \ \mathbf{\theta} )  \ P( \{\hat{\mathbf{\mu}}_i\} |\ \mathbf{\mu}, \ \mathbf{\theta}) \  P(\mathbf{\mu}| \mathbf{\theta}) \  \rm{d}\mathbf{\mu} \ \ ,
\end{equation}

\noindent which is the same up to a constant evidence factor $P( \{\hat{\mathbf{\mu}}_i\})$, as would be expected from equations~\ref{eq:correct_bayes} and~\ref{eq:correct_bayes2}. Which of the previous two distributions one wishes to think of as the traditional {\it likelihood} is somewhat academic, as once they are included into Bayes' theorem (equation ~\ref{eq:correct_bayes} or~\ref{eq:correct_bayes2}), the posterior is the same.

 Adding the corrected likelihood into equation~\ref{eq:correct_bayes} gives the posterior distribution for the parameters of interest

\begin{equation} \label{eq:posterior}
P(\mathbf{\theta} | \mathbf{d}, \{\hat{\mathbf{\mu}}_i \}) = \frac{P(\mathbf{\theta}) }{P(\mathbf{d}) \ P(\{\hat{\mathbf{\mu}}_i\} )} \ \int P(\mathbf{d}\ | \mathbf{\mu} )  \ P( \{\hat{\mathbf{\mu}}_i\} |\ \mathbf{\mu}) \  P(\mathbf{\mu}| \mathbf{\theta}) \  \rm{d}\mathbf{\mu}   \ .
\end{equation}

\noindent Here and in what follows, we drop explicit dependence on $\mathbf{\theta}$ in terms like $P( \{\hat{\mathbf{\mu}}_i\} |\ \mathbf{\mu}, \ \mathbf{\theta})$, and distributions will only condition on the previous layer in the hierarchical model (figure~\ref{fig:network}), as is a common practice (e.g.~\citealt{LeistedtHierarchical}). One can nevertheless keep in mind that probability distributions with $\mathbf{\mu}$ is also conditioned on $\mathbf{\theta}$.

Equation~\ref{eq:posterior} is the most general form, but each factor in the final integral can be evaluated for certain forms of the naive likelihood ($P(\mathbf{d}\ |\ {\mathbf{\mu}})$) and chosen priors:

\begin{itemize}

\item The first factor is the naive likelihood that would be used if the summary statistic $\mathbf{\mu}$ could be calculated.

\item The second factor is a sampling distribution of $\hat{\mathbf{\mu}}$. If our simulated data sets $\mathbf{d}_{{\rm sim}, j}$ for a given set of parameters $\mathbf{\theta}_i$ are independent and realistic, then each is an independent and identically distributed (i.i.d.) draw from the naive likelihood distribution. Assuming we know the naive likelihood, it is usually possible to evaluate the sampling distribution $P(\hat{\mathbf{\mu}}\ |\ \mathbf{\mu})$. 

\item The final factor of equation~\ref{eq:marginal} is the prior on the summary statistic $\mathbf{\mu}$ conditional on position in parameter space. This is not assigning prior probabilities to the values of the parameters themselves but on to possible forms of the summary statistic.

If we believe, for example, that $\mu$ should vary smoothly in parameters space, this could be enforced through a smoothness prior $P(\mathbf{\mu}| \mathbf{\theta})$. This can be compared to emulation methods (e.g.~\citealt{coyoteii},~\citealt{bird_emulator},~\citealt{jennings_emulator}), where $\mathbf{\mu}$ is estimated from simulations using a smoothing prior, either explicitly (Gaussian processes) or implicitly (machine learning). However, the uncertainties are not generally included in the final posterior distribution in a principled hierarchical way as described here.

Conversely, if we claim to know nothing (or very little) about $\mathbf{\mu}$ \textit{a priori}, then we might consider a uniform prior (Section~\ref{sec:known}).

\end{itemize}

\subsection{Bayesian Hierarchical Model}

The model described in the previous section is hierarchical and has a network of parameters related by conditional probabilities. Specific probability distributions of interest, such as the posterior probability distribution of the parameters, are evaluated by appropriate use of Bayes' theorem and marginalization.

The left panel of figure~\ref{fig:network} shows the probabilistic graphical representation of the hierarchical model pertaining to equation~\ref{eq:posterior}. This graphical representation may make the logical steps of the previous section clearer, especially allowing us to see where conditions can be dropped (e.g. $P(\mathbf{d}\ | \mathbf{\mu}, \ \mathbf{\theta})  =  P(\mathbf{d}\ | \mathbf{\mu}) $).

\section{Gaussian Naive Likelihood} \label{sec:gaussian}

For many cosmological analyses, the data $\mathbf{d}$ are assumed to have Gaussian noise and are, therefore, drawn from a Gaussian likelihood. In this section we derive the corrected likelihood for the case with known covariance (Section~\ref{sec:known}) and unknown (estimated) covariance (Section~\ref{sec:unknown}).

A Gaussian likelihood is usually an approximation, as there are likely to be some sources of non-Gaussian noise. It may be a very good approximation however. By the central limit theorem it may be the correct distribution in some limit of large numbers. For example, in a survey to measure the matter power spectrum ${\rm{P}}(k)$, if the galaxies are a Poisson process, then for modes that average many galaxies (high $k$) the likelihood is approximately Gaussian. Similarly, if weak lensing peaks are Poissonian, the binned counts of peaks will be approximately Gaussian for large numbers of observed peaks.

For cases where the naive likelihood is non-Gaussian and one wishes to calculate the corrected likelihood (conditional on an estimated summary statistic $\hat{\mathbf{\mu}}$), one should evaluate equation~\ref{eq:marginal} analytically or numerically.

\subsection{Known Covariance} 
\label{sec:known}

Consider the case where the naive likelihood (the first factor in the integral of equation~\ref{eq:posterior}) is a Gaussian/normal distribution, such that

\begin{equation} \label{eq:gaussian_likelihood}
P(\mathbf {d}\ |\ \mathbf{\mu}(\mathbf{\theta})\ ) = \mathcal{N} (\mathbf {d},\ \mathbf{\mu},\ \mathbf{\Sigma} ) \ \ ,
\end{equation}

\noindent and the covariance $\mathbf{\Sigma}$ is assumed known.

In this case, the sampling distribution (the second factor in the integral of equation~\ref{eq:posterior}) is

\begin{equation} \label{eq:gaussian_sampling}
P (\hat{\mathbf{\mu}}_i\ |\ \mathbf{\mu}_i,\ \mathbf{\Sigma}  ) =  \mathcal{N} (\hat{\mathbf{\mu}}_i,\ \mathbf{\mu}_i,\ \frac{1}{M} \mathbf{\Sigma} ) \ \ ,
\end{equation}

\noindent at position $i$ in parameter space.

If we do not wish to enforce any prior knowledge about $\mathbf{\mu}$, it seems reasonable to use the Jeffreys prior~(\citealt{Jeffreys453}, \citealt{jeffreys1998theory}) as an objective prior distribution for $\mathbf{\mu}$, which, for our Gaussian likelihood (equation~\ref{eq:gaussian_likelihood}), is uniform

\begin{equation} \label{eq:flat_prior}
P(\mathbf{\mu} | \mathbf{\theta}) = P(\mathbf{\mu}) \propto 1 \ \ .
\end{equation}

\noindent This flat prior on $\mathbf{\mu}$, if unbounded, is formally improper. However, the resulting (posterior-like) distribution, $P (\mathbf{\mu}_i\ |\ \hat{\mathbf{\mu}_i},\ \mathbf{\Sigma} )$, is Gaussian and therefore a true probability distribution.

With these distributions (equation~\ref{eq:gaussian_likelihood}-\ref{eq:flat_prior}) for a Gaussian naive likelihood, we perform the marginalization integration (equation~\ref{eq:marginal}),

\begin{equation}
\begin{split}
 P(\mathbf {d}\ |\ \hat{\mathbf{\mu}}_i,\ \Sigma) &= \frac{1}{P(\{\hat{\mathbf{\mu}}_i\})} \int \mathcal{N} (\mathbf{d},\ {\mathbf{\mu}},\ \Sigma) \ \mathcal{N} (\hat{\mathbf{\mu}}_i,\ \mathbf{\mu},\ \frac{1}{M} \mathbf{\Sigma} )\ \rm{d} \mathbf{\mu}  \\
&\propto  \int  \mathcal{N} (\mathbf{d},\ \hat{\mathbf{\mu}}_i, \Big( \frac{M + 1}{M} \Big) \mathbf{\Sigma})\  \mathcal{N}  (\mathbf{\mu},\ \mathbf{x},\ \mathbf{X} ) \ \rm{d} \mathbf{\mu}   \ ,
 \end{split}
\end{equation}

\noindent where $\mathbf{x}$ is a certain function of \{$\mathbf{d}$, $\mathbf{\hat{\mu}}$, $\mathbf{\Sigma}$, $M$\} (but not $\mathbf{\mu}$) and $\mathbf{X}$ is a certain function of  \{$\mathbf{\Sigma}$, $M$\}\footnote{This can be shown by completing the square, and is listed in~\cite{bromiley2003products} and \cite{matrix_cookbook}.}. The first factor can be brought outside the integral. The integration over $\mathbf{\mu}$ of the second factor, which is a normal distribution, evaluates to one, which removes the dependence on $\mathbf{x}$. Normalizing the resulting distribution with respect to $\mathbf{d}$ gives the corrected likelihood:

\begin{equation} \label{eq:corrected_gaussian_known}
P(\mathbf {d}\ |\ \hat{\mathbf{\mu}}_i,\ \Sigma) = \mathcal{N} (\mathbf{d},\ \hat{\mathbf{\mu}}_i,\ \Big( \frac{M + 1}{M} \Big) \mathbf{\Sigma}) \ .
\end{equation}

\noindent For summary statistics ($\hat{\mathbf{\mu}}_i = \hat{\mathbf{\mu}} (\mathbf{\theta}_i)$) estimated from simulations (run with $\mathbf{\theta}_i$ parameters), where the likelihood distribution for data $\mathbf{d}$ conditional on the true (but unknown) $\mathbf{\mu}_i$ is Gaussian with known covariance $\mathbf{\Sigma}$, and with an objective Jeffreys prior on $\mathbf{\mu}$, then equation~\ref{eq:corrected_gaussian_known} is the corrected form of the likelihood. It is this corrected likelihood that should be used for parameter inference.

In this case, the corrected likelihood has the same Gaussian form as the naive likelihood, but with a scaled covariance. At first glance, this scaling could be mistaken for Bessel's correction for an unbiased estimate of the sample variance; however, here we actually know the covariance $\mathbf{\Sigma}$ and the added scaling comes from uncertainty in our estimate $\hat{\mathbf{\mu}}$.
 
\subsection{Unknown Covariance} 
\label{sec:unknown}

In the previous section, we assumed that the summary statistic $\mathbf{\mu}$ was estimated from simulations, but the covariance $\mathbf{\Sigma}$ was known. This situation is unlikely and it is foreseeable that both the summary statistic and covariance would also be estimated from simulated data.

The estimate of the covariance from $N$ independent data simulations is given by
\begin{equation}
\hat{\mathbf{S}} = \frac{1}{N - 1} \sum_{i=0}^{N-1} ( \mathbf{d}_{\rm sim, i} - \bar{\mathbf{d}}_{\rm sim} ) ( \mathbf{d}_{\rm sim, i} - \bar{\mathbf{d}}_{\rm sim} )^\dagger \ ,
\end{equation}

\noindent where $\dagger$ is the conjugate transpose, and $\bar{\mathbf{d}}_{\rm sim} = \frac{1}{N} \sum_{i=0}^{N - 1} \mathbf{d}_{{\rm sim},i}$ is the just the estimated summary statistic $\hat{\mathbf{\mu}}$ as given by equation~\ref{eq:mean_estimate} (but with $N$ simulations, not $M$).

For the case where the summary statistic $\mathbf{\mu}$ is not estimated from simulations, \cite{sh16} calculate the corrected likelihood

\begin{equation} \label{eq:marginal_two}
P(\mathbf {d}\ |  \mathbf{\mu},\ \hat{\mathbf{S}},\ N) \propto \int  P(\mathbf{d}\ |\ {\mathbf{\mu}},\ \mathbf{\Sigma}) P (\hat{\mathbf{S}}\ |\ \mathbf{\Sigma}) P(\mathbf{\Sigma}) \ \rm{d} \mathbf{\Sigma} \ .
\end{equation}

\noindent For a Gaussian naive likelihood $P(\mathbf{d}\ |\ {\mathbf{\mu}},\ \mathbf{\Sigma})$ the distribution of the estimated covariance $P (\hat{\mathbf{S}}\ |\ \mathbf{\Sigma})$ is Wishart. With these distributions and a Jeffreys prior for $\mathbf{\Sigma}$, the resulting Sellentin-Heavens corrected likelihood is given by

\begin{equation} \label{eq:sh_likelihood}
\begin{split}
P(\mathbf {d}\ |\ & \mathbf{\mu},\ \hat{\mathbf{S}}) = \\ &\frac{ \Gamma \big( \frac{N}{2} \big) |\hat{\mathbf{S}}|^{-1/2} } {\Gamma \big( \frac{N-p}{2} \big) \big[ \pi (N - 1)   \big]^{\frac{p}{2}} }\left[1+\frac{(\mathbf{d} - \mathbf{\mu})^\dagger {\hat{\mathbf{S}}}^{-1}(\mathbf{d} - \mathbf{\mu})}{N-1} \right]^{-\frac{N}{2}} \ .
\end{split}
\end{equation}

\noindent where $p$ is the number of elements in the data vector $\mathbf{d}$ (i.e. the dimensionality). This has the form of a multivariate t-distribution.

In the case considered in this work, represented by the right panel of figure~\ref{fig:network} we are assuming that the summary statistic $\mathbf{\mu}$ cannot be calculated, and that we must estimate $\hat{\mathbf{\mu}}$ from simulations.  The integral in equation~\ref{eq:marginal_two} must then be replaced by

\begin{equation} \label{eq:marginal_three}
\begin{split}
P(&\mathbf{d}\ |\{\hat{\mathbf{\mu}}_i\},\  \{\hat{\mathbf{\Sigma}}_i\},\ \mathbf{\theta} ) = \frac{1}{P(\{\hat{\mathbf{\mu}}_i\}) 
 P(\{\hat{\mathbf{\Sigma}}_i\})} \ \times \\ &\iint P(\mathbf{d} | {\mathbf{\mu}},\mathbf{\Sigma}) P (\{\hat{\mathbf{S}}_i\}|\mathbf{\Sigma}) P(\{\hat{\mathbf{\mu}}_i\}|\mathbf{\mu},\mathbf{\Sigma})  P(\mathbf{\Sigma}| \mathbf{\theta})  P(\mathbf{\mu}| \mathbf{\theta}) \  \rm{d}\mathbf{\mu}  \  \rm{d}\mathbf{\Sigma}  \ .
\end{split}
\end{equation}

\noindent This integral is the (corrected) likelihood in the posterior distribution of $\mathbf{\theta}$ for the model shown in the right-hand panel of figure~\ref{fig:network}.

We note that a given sample mean $\hat{\mathbf{\mu}}_i$ and sample covariance $\hat{\mathbf{S}}_i$ are independent, even if the simulated data used to estimate them are the same~\citep{anderson2003introduction}, so that $P(\hat{\mathbf{\mu}}, \hat{\mathbf{S}}) = P(\hat{\mathbf{\mu}})\ P( \hat{\mathbf{S}})$. If the simulated data used to evaluate $\hat{\mathbf{\mu}}$ and $\hat{\mathbf{S}}$ are the same, one can set $N = M$ in what follows. If, as is often the case in cosmological analyses, the covariance is assumed not to vary with respect to the parameters of interest and is instead estimated once with more simulated data realizations, then $N$ is fixed and the $i$ indices on the estimated covariance are dropped.

Using the same distributions as described so far in Section~\ref{sec:gaussian}, marginalizing over the unknown true summary statistic $\mathbf{\mu}$ and covariance $\mathbf{\Sigma}$ (equation~\ref{eq:marginal_three}) and renormalizing gives the new corrected likelihood (details in Appendix~\ref{full_derivation}):

\begin{equation} \label{eq:new_likelihood}
\begin{split}
&P(\mathbf {d}\ |\ \hat{\mathbf{\mu}}_i,\ \hat{\mathbf{S}}_i) = \\ &\frac{ \Gamma \big( \frac{N}{2} \big) \sqrt[]{ \frac{M}{M+1} } |\hat{\mathbf{S}}_i|^{-1/2} } {\Gamma \big( \frac{N-p}{2} \big) \big[ \pi (N - 1)   \big]^{\frac{p}{2}} }\left[1+\frac{M}{(M+1)(N-1)}(\mathbf{d} - \mathbf{\mu}_i)^\dagger {\hat{\mathbf{S}}}_i^{-1}(\mathbf{d} - \mathbf{\mu}_i)\right]^{-\frac{N}{2}} \ .
\end{split}
\end{equation}

\noindent This corrected likelihood gives the probability of observing the data $\mathbf{d}$ conditional on an estimated mean summary statistic $\hat{\mathbf{\mu}}_i=\hat{\mathbf{\mu}}(\mathbf{\theta}_i)$ from $M$ simulations and an estimated covariance matrix $\hat{\mathbf{\Sigma}}_i = \hat{\mathbf{\Sigma}}(\mathbf{\theta}_i)$ from $N$ simulations, where we have assumed that the naive likelihood is Gaussian and have used Jeffreys priors on $\mathbf{\mu}$ and $\mathbf{\Sigma}$.

\section{Toy Model Demonstration} 
\label{sec:toy_model}

As a verification and demonstration of the result given in equation~\ref{eq:corrected_gaussian_known}, where $\hat{\mathbf{\mu}}$ is estimated from simulated data and $\mathbf{\Sigma}$ is known, we construct a toy model. This toy model also relies on the assumed flat prior on $\mathbf{\mu}$ and the fact that the sampling distribution is symmetric with respect to $\mathbf{\mu}$ and $\mathbf{\hat{\mu}}$.

Let us assume that in different realizations of an experiment, different experimenters randomly and independently generate $M$ simulations, from which $\hat{\mathbf{\mu}}$ is estimated according to equation~\ref{eq:mean_estimate}. The underlying likelihood distribution for the data with known $\mathbf{\mu}$ is Gaussian and therefore the simulated data are themselves i.i.d. draws from the Gaussian distribution (equation~\ref{eq:gaussian_likelihood}). Each experimenter then draws a realization of the data $\mathbf{d}$ from the naive Gaussian likelihood distribution with the known variance and the mean given by their estimated $\hat{\mathbf{\mu}}$.

\begin{figure}
\hspace*{-0.05in}
\includegraphics[width=0.47\textwidth]{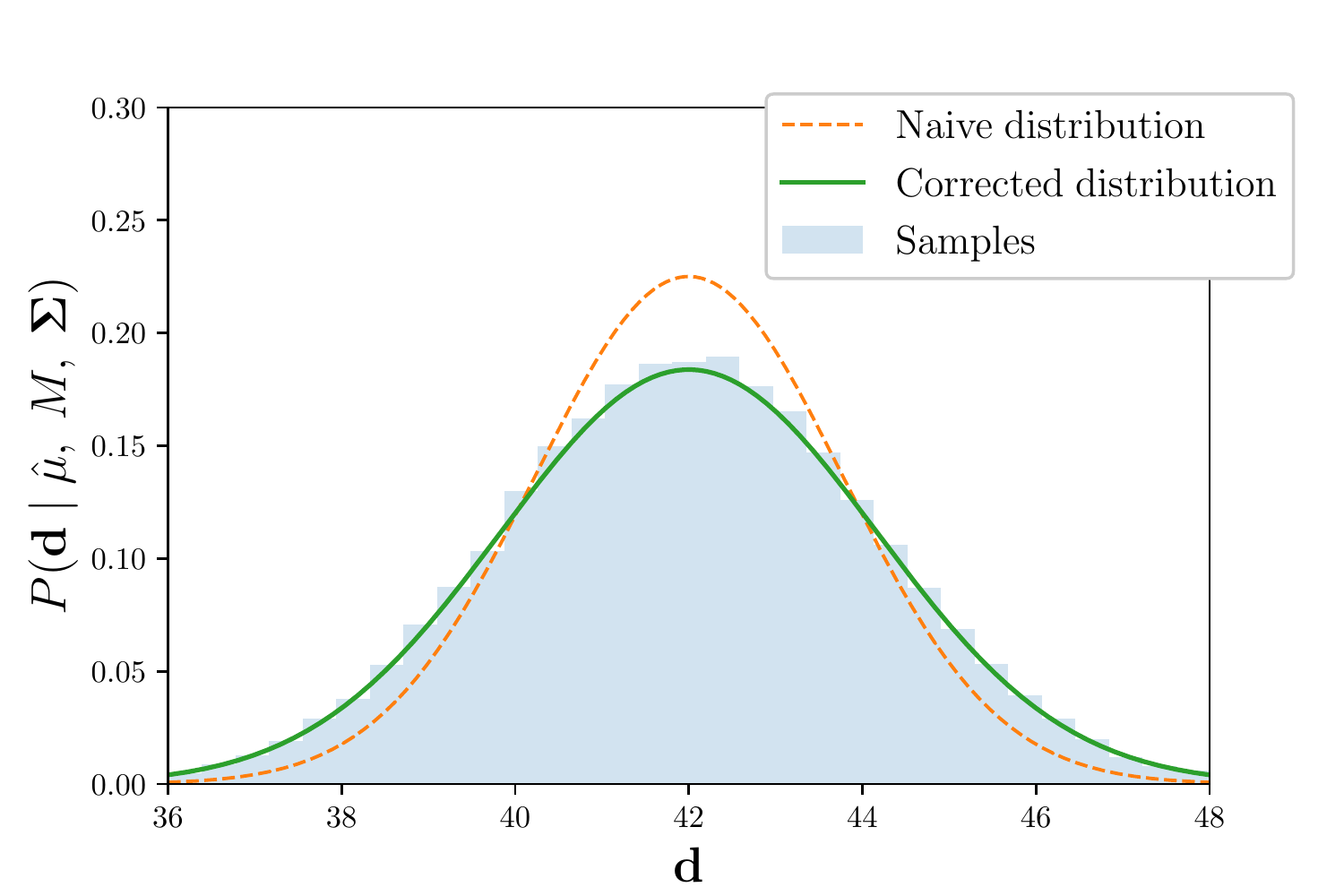}
\caption{\label{fig:distribution} One-dimensional toy model for the naive Gaussian likelihood with known variance, where $\mathbf{\mu} = 42$, $\mathbf{\Sigma} = \pi \approx 3.14 $, and $M=2$. The samples (described in Section~\ref{sec:toy_model}) are distributed according to the corrected likelihood distribution (equation~\ref{eq:corrected_gaussian_known}), whereas the naive likelihood distribution has reduced variance.}
\end{figure} 

\begin{figure*}
\centering
\hspace*{-0.05in}
\vspace*{-.1in}
\includegraphics[width=0.97\textwidth]{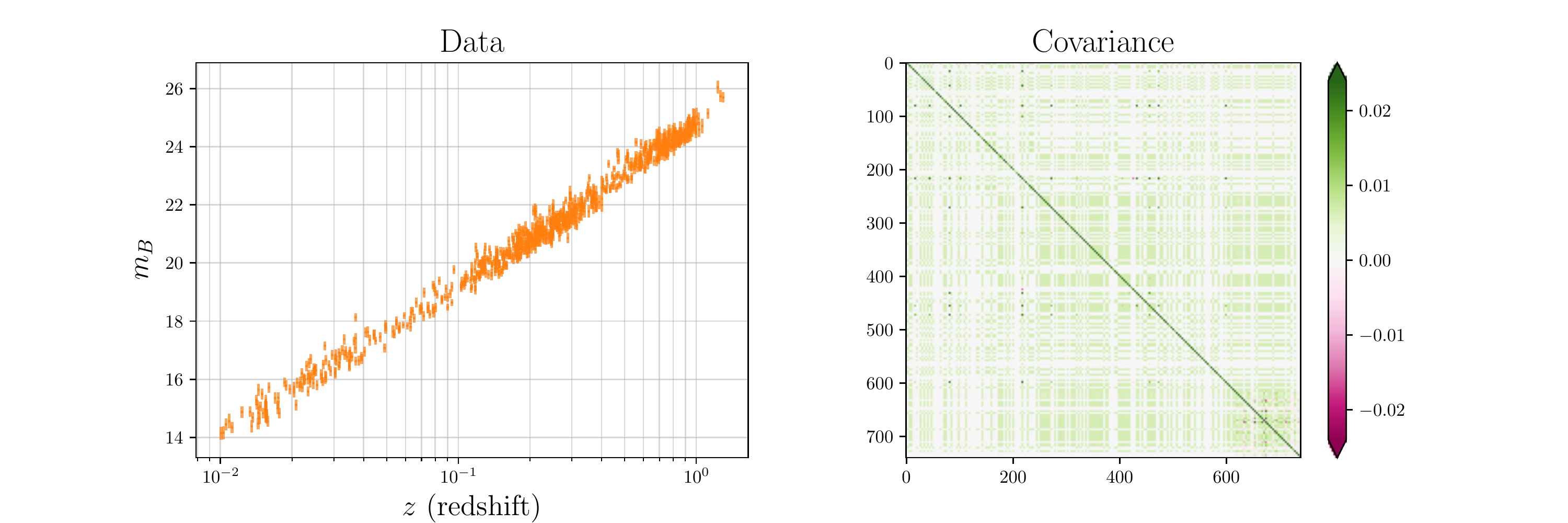}
\caption{\label{fig:data} {\it{Left panel}}: The observed magnitude $m_B$ data for 740 SN Ia. The error bars are taken as the square-root of the diagonal elements of the covariance. {\it{Right panel}}: The covariance matrix as described in Section~\ref{sec:model}.}
\end{figure*} 

Though each experimenter draws their data realization $\mathbf{d}$ from a Gaussian likelihood with the known variance $\mathbf{\Sigma}$, the data realizations from all the experimenters will be distributed according to the corrected likelihood (equation~\ref{eq:corrected_gaussian_known}) with variance $\frac{M + 1}{M} \mathbf{\Sigma}$. 

In figure~\ref{fig:distribution} we take a one-dimensional case where $\mathbf{\mu} = 42$, $\mathbf{\Sigma} = \pi \approx 3.14 $, and $M=2$. We see that the data samples from $10^5$ different experiments match the corrected likelihood distribution (equation~\ref{eq:corrected_gaussian_known}), whereas the naive likelihood distribution underestimates the variance, as expected. 

\section{JLA Supernovae Demonstration} 
\label{sec:JLA}

In this section, we use public type Ia supernova (SN Ia) data\footnote{\url{supernovae.in2p3.fr/sdss_snls_jla/ReadMe.html}} from the SDSS-II/SNLS3 Joint Light-Curve Analysis (JLA)~\citep{jla} as a demonstration of the corrected likelihoods described in the previous sections.

This is, of course, only a demonstration, as the summary statistic $\mathbf{\mu}(\mathbf{\theta})$ (SN Ia apparent magnitudes) can actually be calculated for the model considered. We generate simulated data by drawing realizations from the known likelihood\footnote{\cite{alsing_likelihood_free} and \cite{leclercq_likelihood_free} take  a similar approach to demonstrate likelihood-free methods.}, and estimate $\hat{\mathbf{\mu}}$. We can then constrain cosmological parameters using a likelihood distribution conditional on our estimated $\hat{\mathbf{\mu}}$.

\subsection{Data and Model} \label{sec:model}

The data are observed B-band peak apparent magnitudes $\mathbf{d} = (m_{B,\rm{obs},1}, m_{B,\rm{obs},2}, ... )$ for $740$ SN Ia over a range of redshifts up to $z = 1.3$. The supernovae also have associated light-curve stretch $X_1$, colour at maximum-brightness $C$ and host stellar mass $M_{\rm{stellar}}$, which are included in the model and covariance. The data and associated covariance are shown in figure~\ref{fig:data}. 

We use the model from~\cite{jla} where the SN Ia are standardizable candles with expected apparent magnitude

\begin{equation} \label{eq:model}
\begin{split}
m_B = 5 &\log_{10} \Big( \frac{D_L}{1 \rm{Mpc}}\Big) + 25\\ &+ M_B + \Delta M \Theta (M_{\rm{stellar}} - 10^{10} M_{\sun}) \\ &- \alpha X_1 + \beta C \ ,
\end{split}
\end{equation}

\noindent where $\alpha$ and $\beta$ are nuisance parameters for the stretch and
colour respectively. $M_B$ is the absolute magnitude of the host with a correction term $\Delta M$ depending on $M_{\rm{stellar}}$ (where $\Theta$ is the Heaviside function). We take a flat $w$CDM Universe, with luminosity distance

\begin{equation}
D_L = \frac{c (1+z)}{H_0} \int^z_0  \frac{{\rm{d}} z'}{\sqrt[]{ \Omega_m (1 + z')^3 + (1 - \Omega_m) (1 + z')^{3(w + 1)} } } \ \ ,
\end{equation}

\noindent where $c$ is the speed-of-light in a vacuum, $H_0$ is the Hubble parameter, $\Omega_m$ is the matter density parameter, and $w$ is the equation of state for dark energy.

In our demonstration, the parameters of interest $\mathbf{\theta} = (\Omega_m, w, M_B)$ are allowed to vary. For simplicity, we fix the other parameters: $H_0 = 70.0\ \rm{km}\ \rm{s}^{-1}\ \rm{Mpc}^{-1}$, $\Delta M = -0.05$, $\alpha = 0.1256$, $\beta = 2.6342$. The covariance is calculated according to~\cite{jla} and is also fixed.

\subsection{Likelihood and Priors} \label{sec:likelihood_and_prior}

We assume a Gaussian likelihood, where the log-likelihood is given by:

\begin{equation}
\rm{ln} P(\mathbf{d} | \mathbf{\mu}, \mathbf{\Sigma}) = -\frac{1}{2} \rm{ln} \big[ \rm{det}\big( 2 \pi \mathbf{\Sigma} \big) \big] - \frac{1}{2} (\mathbf{d} - \mathbf{\mu})^\dagger \mathbf{\Sigma}^{-1} (\mathbf{d} - \mathbf{\mu}) \ \ ,
\end{equation}

\noindent where the data and covariance are those described in Section~\ref{sec:model} (and shown in Fig.~\ref{fig:data}), and our summary statistic $\mathbf{\mu}$ is given in equation~\ref{eq:model}.

For simplicity we take uniform priors in the ranges: $0.05 \leq \Omega_m \leq 0.6$, $-1.5 \leq w \leq -0.3$, $-19.15 \leq M_B \leq -18.95$.

Simulations for this demonstration are run on a regular grid of shape $[12, 13, 11]$ (for $\Omega_m$, $w$, $M_B$) spanning the prior range. The regular grid is a particularly poor choice to sample the posterior distribution when simulations are expensive. However, this is a demonstration, and for real-world analysis many better sampling schemes are available (including latin hypercubes and grid transformations to better sample the expected posterior distribution).

Once the posterior is evaluated at these grid positions, the parameter space is upsampled to a $[48, 52, 44]$ grid. The new grid positions are evaluated by interpolating the posterior distribution from the original grid using a radial basis function `thin plate' spline (\citealt{duchon1976interpolation}; \citealt{bookstein1989principal}; \citealt{scipy}). This spline interpolation worked particularly well in avoiding edge effects or artefacts around points when we compared their results with those of more poorly-performing simple polynomial splines.

\begin{figure}
\hspace*{-0.05in}
\vspace*{-.15in}
\includegraphics[width=0.47\textwidth]{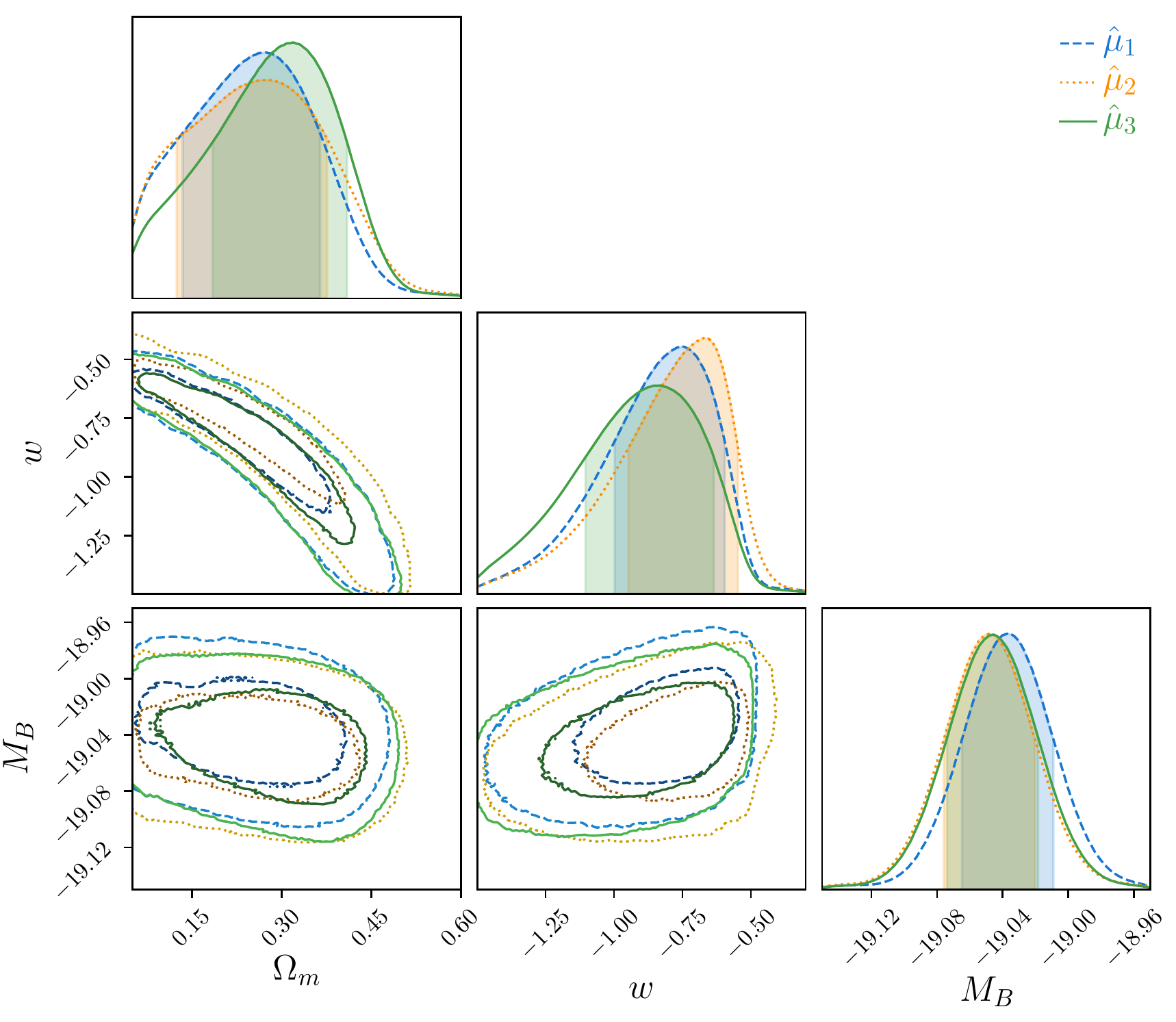}
\caption{\label{fig:different_mu_hat} JLA posterior distribution for $\Omega_m$, $w$, and $M_B$ (described in Section~\ref{sec:likelihood_and_prior}) using three independent estimates $\hat{\mathbf{\mu}}$ with $M=2$ simulations per position in parameter space. This uses the naive Gaussian distribution (equation~\ref{eq:gaussian_likelihood}) without the correction (equation~\ref{eq:corrected_gaussian_known}), and the contours are therefore optimistically reduced.}
\end{figure} 

\subsection{Results}
\label{sec:jla_result}

First, let us imagine three different experimenters, who, despite having access to the same data (described above), run their own independent simulations to estimate the summary statistics $\hat{\mathbf{\mu}}(\mathbf{\theta})$ on the grid in parameter space. This results in different $\hat{\mathbf{\mu}}_i$ for $i =$1, 2, 3. 

Evaluating the posterior distribution using the naive Gaussian likelihood set-up described in Section~\ref{sec:likelihood_and_prior}, and using $M=2$ simulations per parameter grid position, results in the three posterior distributions in figure~\ref{fig:different_mu_hat}. The three different experimenters have three different posterior distributions due to their different $\hat{\mathbf{\mu}}_i$. 

Having different individual posterior distributions is in itself is not a problem. If different experiments have different data but the same underlying parameters, their resulting posterior distributions will look different, and will quantify their own individual uncertainty in the parameters. However, this variance of the data has been taken into account, and will be reflected in each posterior distribution. In the case of different $\hat{\mathbf{\mu}}_i$ in figure~\ref{fig:different_mu_hat}, the fact that $\hat{\mathbf{\mu}}$ was a random draw from a distribution (just like the data) has not been taken into account. As they have ignored the resulting correction to their likelihood, each experimenter will be overly optimistic about their own inferences.

\begin{figure}
\hspace*{-0.05in}
\vspace*{-.15in}
\includegraphics[width=0.47\textwidth]{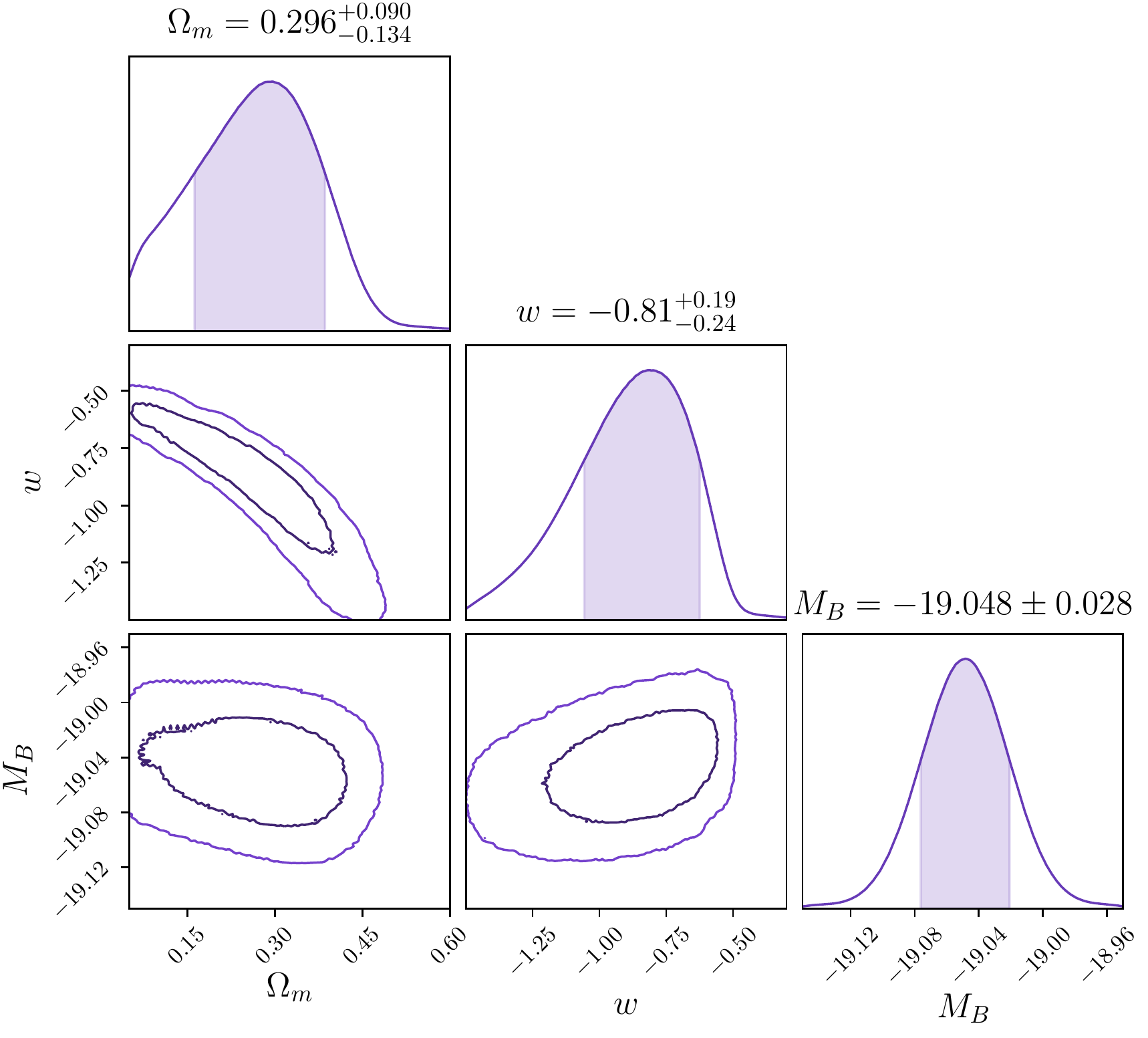}
\caption{\label{fig:corrected_jla} JLA posterior distribution for $\Omega_m$, $w$, and $M_B$ (described in Section~\ref{sec:jla_result}) using the corrected likelihood (equation~\ref{eq:corrected_gaussian_known}) with $M=2$ (we set $\hat{\mathbf{\mu}} = \mathbf{\mu}$ in this example for clarity). The contours are broader than in figure \ref{fig:different_mu_hat} (25 per cent increase in the parameter covariance determinant) as this likelihood takes into account that the estimated summary statistic is a draw from a sampling distribution.}
\end{figure}

In figure~\ref{fig:corrected_jla}, the posterior distribution has been calculated using the likelihood correction (equation~\ref{eq:corrected_gaussian_known}), which takes into account the variance in $\hat{\mathbf{\mu}}_i$ (using the $\mathbf{\mu}$ summary statistic for clarity).  Using ChainConsumer \citep{Hinton2016}, we measure a $25$ per cent increase in the determinant of the parameter covariance with the corrected likelihood. The resulting posterior distribution is  slightly broader, reflecting the added uncertainty in the inference.

\subsection{Model Comparison} \label{sec:model_comparison}

The comparison of different theoretical models using the data in a Bayesian framework is usually done by calculating the Bayesian evidence:

\begin{equation} \label{eq:evidence}
P(\mathbf{d} | {\rm Model}) = \mathcal{Z} = \int P(\mathbf{d} | \mathbf{\theta},\ {\rm Model})\ P( \mathbf{\theta} |\ {\rm Model})\ {\rm{d}} \mathbf{\theta} \ .
\end{equation}

\noindent Two models can be compared by evaluating the Bayes factor:

\begin{equation} \label{eq:bayes_factor}
K = \frac{\mathcal{Z}_1}{\mathcal{Z}_2} = \frac{P(\mathbf{d} | {\rm Model}_1)}{P(\mathbf{d} | {\rm Model}_2)} = \frac{P({\rm Model}_1 | \mathbf{d})}{P({\rm Model}_2 | \mathbf{d})}  \frac{P({\rm Model}_2)}{P({\rm Model}_1)} \ .
\end{equation}

\noindent If one has no reason to believe a given model more than another {\it a priori}, then the second factor (the ratio of the prior distributions) equals one. In this case, the Bayes factor becomes a ratio of the model probabilities (conditional on the data). 

We evaluate the Bayesian evidence (equation~\ref{eq:evidence}) for the three parameter case for the JLA analysis described in Section~\ref{sec:likelihood_and_prior} with the uncorrected naive Gaussian likelihood and the corrected likelihood (equation~\ref{eq:corrected_gaussian_known}) with $M=2$. Calculating the Bayes factor gives

\begin{equation} \label{eq:bayes_factor_results}
\begin{split}
{\rm ln } K_{\hat{\mu}_1}  &= 46 \\
{\rm ln }K_{\hat{\mu}_2}&= 31\ \\
{\rm ln }K_{\hat{\mu}_3}&= 30\ ,
\end{split}
\end{equation}

\noindent for $\hat{\mathbf{\mu}}_1$, $\hat{\mathbf{\mu}}_2$, and $\hat{\mathbf{\mu}}_3$ respectively, where

\begin{equation}
{\rm ln } K = {\rm ln } \mathcal{Z}_{\rm{\scriptstyle{ corrected}}} - {\rm ln }  \mathcal{Z}_{\rm{\scriptstyle{uncorrected}}}
\end{equation}

\noindent using the corrected likelihood to evaluate the evidence $\mathcal{Z}_{\rm{\scriptstyle{corrected}}}$ and the naive, uncorrected likelihood for $\mathcal{Z}_{\rm{\scriptstyle{uncorrected}}}$.

As a check, after increasing the number of evaluated grid points by a factor of 4 we still calculate the same $K$ values. Additionally, we calculate ${\rm ln}K$ using a different cosmological parametrisation, sampling scheme and data (Appendix~\ref{sec:pantheon}) and get similar results.

For all three, the corrected likelihood is more than a factor of ${\rm{exp}}[30]$ more probable than the uncorrected. This is further validation of the corrected likelihood; the model (i.e. the corrected likelihood) shows a better goodness-of-fit. Furthermore, if one were using an estimated summary statistic, but not using the corrected likelihood, one's belief in a model would be incorrect by this factor. 

This effect would be particularly harmful if comparing two models, where it is possible to calculate $\mathbf{\mu}$ for the first, but $\hat{\mathbf{\mu}}$ is estimated from simulations for the second. Using the same Gaussian likelihood for both, without the correction for the second, would lead one to incorrectly favour the first model.

\begin{figure}
\hspace*{-0.in}
\vspace*{-.15in}
\includegraphics[width=0.45\textwidth]{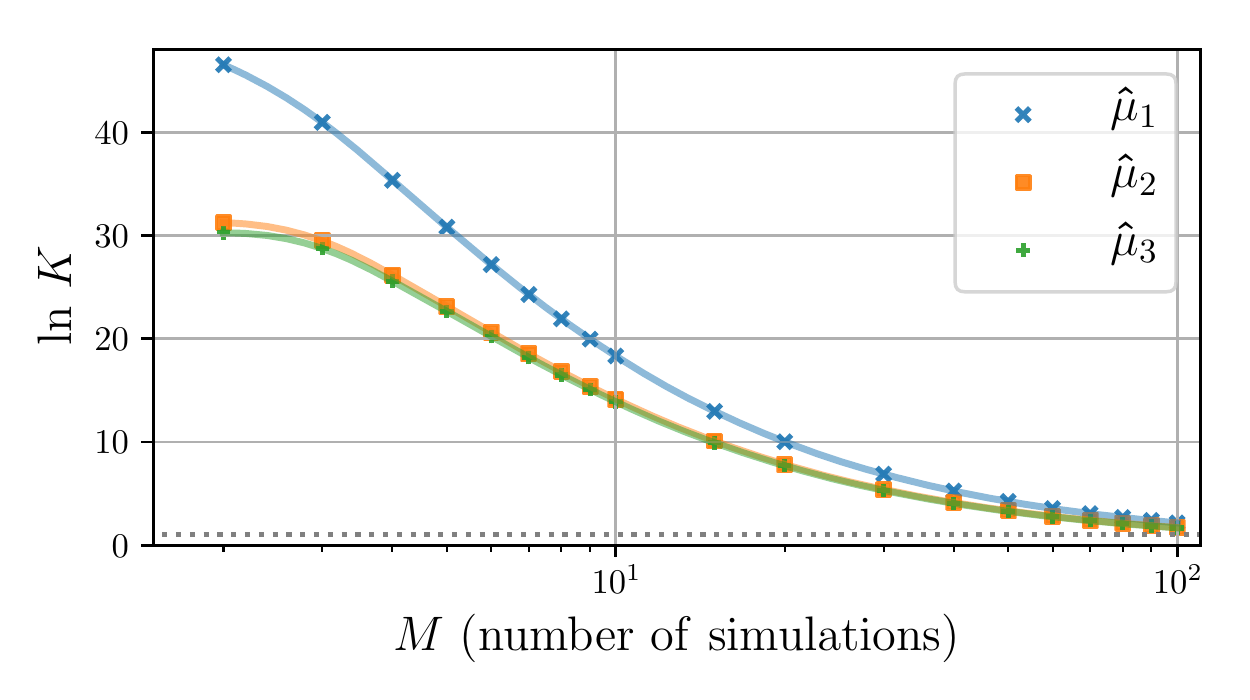}
\caption{\label{fig:evidence} The log Bayes factor $\ln K$ as a function of the number of simulations $M$ for the three estimated summary statistics $\hat{\mathbf{\mu}}_i$ (using cubic spline interpolation between evaluated points).}
\end{figure}

In figure~\ref{fig:evidence}, the log Bayes factors $\ln K$ (equation~\ref{eq:bayes_factor_results}) are shown for the three estimated summary statistics $\hat{\mathbf{\mu}}_i$ as a function of the number of simulations $M$. As the number of simulations increase, the error incurred by using the uncorrected likelihood decreases. The value asymptotes to, but will never reach, $\ln K = 0$. For even a large number of simulations (large $M$), the error is not negligible. For many analyses, rather than running $10^2$ expensive simulations, it would be better to use the corrected likelihood and avoid this error.

\section{Discussion \& Conclusions}

In this work, we have shown how to take the sampling distribution of estimated summary statistics, $\hat{\mathbf{\mu}}$, into account for parameter inference in a cosmological context. For situations where the naive likelihood is Gaussian, we have evaluated this correction (by marginalizing over the unknown $\mathbf{\mu}$) for the case with known covariance (equation~\ref{eq:corrected_gaussian_known}) and estimated covariance (equation~\ref{eq:new_likelihood}).

We have validated the corrected likelihood with a toy model (Section~\ref{sec:toy_model}). Using JLA SN Ia data, we have demonstrated the effect of the corrected likelihood on cosmological parameter inference. For model selection, in our simple three-parameter inference demonstration, we show that the log Bayesian evidence ${\rm ln} \mathcal{Z}$ will be incorrect with a difference of over $30$ if the uncorrected likelihood is not used.

In the era of DES~\citep{des_y1_results}, LSST~\citep{lsst_science} and Euclid~\citep{euclid_science}, cosmological analyses will have access to large cosmological data sets. Sole reliance on two-point statistics in the linear regime will be tantamount to wasting data which is rich in cosmological information. However, many summary statistics ($\mathbf{\mu}$) that access information beyond these two-point statistics in the linear regime cannot be calculated analytically and need realistic simulations to be run to estimate $\hat{\mathbf{\mu}}$.

A typical approach has been to run an excessive number of simulations at each position in parameter space, such that the variance of $\hat{\mathbf{\mu}}$ in negligible. This approach has diminishing returns, as variance asymptotically tends to zero as $\frac{1}{M}$. This effectively aims to increase the number of simulations $M$ until the sampling distribution $P (\hat{\mathbf{\mu}}\ |\ \mathbf{\mu})$ can be effectively viewed as Dirac delta function (a limit it will, of course, never reach).  Accepting a small increase in the resulting parameter constraints and correcting the likelihood for this sampling distribution means that fewer simulations have to be run.

If one does not wish to take the sampling distribution into account, one might use ``cheap'' simulations where it is possible to run enough that one effectively reaches the Dirac delta function limit. This has two potential pitfalls: firstly, the limit is never truly reached, which may affect the inferred parameters or model comparison results; secondly, cheap simulations are likely to be less realistic. It is far better to have slightly broader posterior distributions and to have used reliable simulations, than to have tighter constrains on parameters that are biased due to unreliable simulations.

The approach taken in this paper requires the acceptance that simulations are not ``free''. Simulations are increasingly an essential part of analyses. Like data, reliable simulations are often expensive in terms of time and resources and are, therefore, an acceptable contribution to the uncertainty of inferred parameters.

\section*{Acknowledgements}

We thank Alan Heavens, Ofer Lahav, Pablo Lemos, Konstantinos Themelis and Lorne Whiteway for extremely useful comments and discussions, and Andrei Cuceu for his assistance with {\sc Pliny}. This work is funded by the DEDALE project, contract no. 665044, within the H2020 Framework Program of the European Commission. NJ also acknowledges the UK Science and Technology Research Council (STFC) Grant No. ST/M001334/1.

%%%%%%%%%%%%%%%%%%%%%%%%%%%%%%%%%%%%%%%%%%%%%%%%%%



\bibliographystyle{mnras}
\bibliography{bibliog} % if your bibtex file is called example.bib

%%%%%%%%%%%%%%%%%%%%%%%%%%%%%%%%%%%%%%%%%%%%%%%%%%

%%%%%%%%%%%%%%%%% APPENDICES %%%%%%%%%%%%%%%%%%%%%

\appendix

\section{} \label{full_derivation}

Here the full derivation leading to equation~\ref{eq:new_likelihood} is given, starting from the integration equation~\ref{eq:marginal_three},

\begin{equation} 
\begin{split}
P(&\mathbf{d}|\{\hat{\mathbf{\mu}}_i\},\ \{\hat{\mathbf{S}}_i\}) \propto  \\ &\iint  P(\mathbf{d} | {\mathbf{\mu}},\mathbf{\Sigma}) P (\{\hat{\mathbf{S}}_i\}|\mathbf{\Sigma}) P(\{\hat{\mathbf{\mu}}_i\}|\mathbf{\mu},\mathbf{\Sigma}) P(\mathbf{\Sigma}|\mathbf{\theta}) P(\mathbf{\mu} | \mathbf{\theta})\ \rm{d} \mathbf{\Sigma}\ \rm{d} \mathbf{\mu}   \\
&\int  \Bigg[\int P(\mathbf{d} | {\mathbf{\mu}},\mathbf{\Sigma}) P(\{\hat{\mathbf{\mu}}_i\}|\mathbf{\mu},\mathbf{\Sigma})  P(\mathbf{\mu}| \mathbf{\theta}) \ {\rm{d}} \mathbf{\mu}  \Bigg] P (\{\hat{\mathbf{S}}_i\}|\mathbf{\Sigma}) P(\mathbf{\Sigma}| \mathbf{\theta}) \rm{d} \mathbf{\Sigma}   \ \ .
\end{split}
\end{equation}

\noindent The first factor in the integral, $P(\mathbf{d} | {\mathbf{\mu}},\mathbf{\Sigma})$, is the Gaussian naive likelihood, given by equation~\ref{eq:gaussian_likelihood}. The final factor in the first integral is a uniform prior, $P(\mathbf{\mu}) \propto 1$, as described in Section~\ref{sec:known}. The second line uses the fact that $\hat{\mathbf{\mu}}$ and $\hat{\mathbf{S}}$ are independent~\citep{anderson2003introduction}. We can first perform the integration over $\mathbf{\mu}$, which, using the result from Section~\ref{sec:known}, gives

\begin{equation} \label{eq:first_integration}
P(\mathbf {d}\ |\hat{\mathbf{\mu}}_i,\ \hat{\mathbf{S}}_i) \propto  \int  \mathcal{N} (\mathbf{d}, \hat{\mathbf{\mu}}_i, \Big( \frac{M + 1}{M} \Big) \mathbf{\Sigma}) P (\hat{\mathbf{S}}_i|\mathbf{\Sigma}) P(\mathbf{\Sigma} | \mathbf{\theta}) \rm{d} \mathbf{\Sigma}  \ .
\end{equation}

\noindent The second factor is the Wishart distribution

\begin{equation}
\begin{split}
P(\hat{\mathbf{S}}_i|\mathbf{\Sigma},N) &=  \mathcal{W}(\hat{\mathbf{S}}_i | \mathbf{\Sigma}/(N-1), N-1 ) \\&=  \frac{|\hat{\mathbf{S}}_i|^\frac{N-p-2}{2} \exp\left[ - \frac{N-1}{2} \rm{Tr} \big[{\mathbf{\Sigma}^{-1} \hat{\mathbf{S}}_i } \big] \right]    }{2^\frac{p(N-1)}{2} |\mathbf{\Sigma}/(N-1)|^\frac{N - 1}{2}  \Gamma_p\left( \frac{N - 1}{2} \right)} \ \ ,
\end{split}
\end{equation}

\noindent and the third factor is the Jeffreys prior for $\mathbf{\Sigma}$

\begin{equation}
P(\mathbf{\Sigma}) \propto | \mathbf{\Sigma} |^{-\frac{p+1}{2}} \ \ ,
\end{equation}

\noindent as described by~\cite{sh16}. 

With these factors, we can rewrite equation~\ref{eq:first_integration}, to give

\begin{equation} 
\begin{split}
P(\mathbf {d}&\ | \hat{\mathbf{\mu}}_i,\ \hat{\mathbf{S}}_i) \\ &\propto \int  \rm{d} \mathbf{\Sigma}\  | \mathbf{\Sigma} |^{-\frac{N+p+1}{2}} \ \exp \Big[ - \frac{1}{2} \rm{Tr} \big( \mathbf{\Sigma}^{-1} \mathbf{\phi}_i \big) \Big] \\
&\propto  |\mathbf{\phi}_i|^{-N/2} \ , 
\end{split}
\end{equation}

\noindent where

\begin{equation}
\mathbf{\phi}_i = (N - 1)\  \hat{\mathbf{S}}_i + \mathbf{v}_i \mathbf{v}_i^\dagger \ \ ,
\end{equation}

\noindent and

\begin{equation}
\mathbf{v}_i = (\mathbf{d} - \hat{\mathbf{\mu}}_i ) \ \sqrt[]{\frac{M}{M + 1}} \ \ .
\end{equation}

\noindent Using the identity $|\mathbf{A} + \mathbf{v} \mathbf{v}^\dagger| = |\mathbf{A}||1 + \mathbf{v}^\dagger \mathbf{A}^{-1} \mathbf{v}|$ and normalizing gives the new corrected likelihood in equation~\ref{eq:new_likelihood}. The result is discussed in Section~\ref{sec:unknown}.

\section{} \label{sec:pantheon}

In Section~\ref{sec:model_comparison}, the Bayes factors were calculated by evaluating the integral (equation~\ref{eq:evidence}) numerically on the parameter grid described in Section~\ref{sec:likelihood_and_prior}. This resulted in values of $\ln K > 30$ for the corrected vs. uncorrected likelihood using the JLA data with the parameters described. Here we briefly describe how we verified the magnitude of this effect using 1048 Pantheon SN Ia~\citep{pantheon} with a different set of parameters and a different method to evaluate the Bayesian evidence.

We allow four parameters to vary: $q_0$ (deceleration), $j_0$ (jerk), $M_B$, and $h = H_0/ [100.0\ \rm{km}\ \rm{s}^{-1}\ \rm{Mpc}^{-1}]$. Deceleration and jerk are the parameterisation of a third-order Taylor expansion of the scale factor $a(t)$~\citep{jerk_snap}. The priors are uniform in the ranges: $-1<q_0<1$, $-2<j_0<2$, $-20<M_B<-18$, and $0.4<h<1.2$.  To evaluate the posterior distribution and Bayesian evidence we use {\sc Pliny}~\citep{pliny}, which is a nested sampler~\citep{nested_sampling}, and has been shown to be accurate when compared with known closed-form Bayesian evidence results.

The Bayes factor $K$ is also differently defined here, as we set $\mathbf{\mu} = \hat{\mathbf{\mu}}$, inverting the result. The evaluated value $\ln K \approx -{43}$, validates our results from Section~\ref{sec:model_comparison}.

%%%%%%%%%%%%%%%%%%%%%%%%%%%%%%%%%%%%%%%%%%%%%%%%%%

% Don't change these lines
\bsp	% typesetting comment
\label{lastpage}
\end{document}